\documentclass[aps,nofootinbib,preprint,showpacs,preprintnumbers,amsmath,amssymb]{revtex4}%
\usepackage{latexsym}%
\usepackage{epsf}%
\usepackage[hypertex]{hyperref}%

\def\cF{{\mathcal F}}

\def\cW{{\mathcal W}}

\def\acoth{{\text{Arccoth}}}
\def\atanh{{\text{Arctanh}}}

\newcommand{\beq}{\begin{equation}}
\newcommand{\beqn}{\begin{equation}\nonumber}
\newcommand{\eeq}{\end{equation}}
\newcommand{\bea}{\begin{eqnarray}}
\newcommand{\bean}{\begin{eqnarray}\nonumber}
\newcommand{\eea}{\end{eqnarray}}

\begin{document}

\begin{center}
{\bf{\Large Reflection and Transmission at the Apparent Horizon during Gravitational Collapse}}
\bigskip
\bigskip

{{Cenalo Vaz$^{a,b,}$\footnote{e-mail address: Cenalo.Vaz@UC.Edu},
L.C.R. Wijewardhana$^{b,}$\footnote{e-mail address: Rohana.Wijewardhana@UC.Edu}}}
\bigskip

{\it$^a$RWC and $^b$Department of Physics,}\\
{\it University of Cincinnati,}\\
{\it Cincinnati, Ohio 45221-0011, USA}
\end{center}
\bigskip
\bigskip
\medskip

\centerline{ABSTRACT}
\bigskip\bigskip

We examine the wave-functionals describing the collapse of a self-gravitating dust ball in an 
exact quantization of the gravity-dust system. We show that ingoing (collapsing) dust shell modes 
outside the apparent horizon must necessarily be accompanied by outgoing modes inside the 
apparent horizon, whose amplitude is suppressed by the square root of the Boltzmann factor at 
the Hawking temperature. Likewise, ingoing modes in the interior must be accompanied by 
outgoing modes in the exterior, again with an amplitude suppressed by the same factor. A 
suitable superposition of the two solutions is necessary to conserve the dust probability flux 
across the apparent horizon, thus each region contains both ingoing and outgoing dust modes. 
If one restricts oneself to considering only the modes outside the apparent horizon then one 
should think of the apparent horizon as a partial reflector, the probability for a shell to reflect 
being given by the Boltzmann factor at the Hawking temperature determined by the mass 
contained within it. However, if one considers the entire wave function, the outgoing wave 
in the exterior is seen to be the transmission through the horizon of the interior outgoing wave 
that accompanies the collapsing shells. This transmission could allow information from the interior to 
be transferred to the exterior.
\vfill\eject

\section{Introduction}

Hawking's derivation of black hole radiance \cite{haw75} was given within a field 
theoretic context, by calculating the Bogliubov transformation of scalar field operators in 
the Schwarzschild spacetime. The physical process responsible for the radiation was,
however, generally taken to be a quasi-classical tunneling of particles through the horizon. 
In the tunneling picture, pair production occurs near the horizon. A negative energy 
antiparticle falls in and a positive energy particle is radiated out, leading to a decrease
in the black hole mass. Thus the intuitive picture and the actual computation were not 
directly comparable until, many years later, the tunneling picture was directly used in 
\cite{pw00,p04} to provide a new semi-classical derivation of the Hawking effect (see also 
\cite{kw99,sp99,v99}).  Developments and refinements of these methods can be found in 
\cite{amv05,mv05,a06,bm08}. 

In the tunneling picture one computes the tunneling rate as the exponential of the imaginary part 
of the classical action for scalar particles in a neighborhood of the horizon. This formalism has 
also been applied to de Sitter space in \cite{vol08} and shown to have connections with black hole 
thermodynamics in \cite{rb08}. Another approach to understanding Hawking radiation is via 
the gravitational anomalies of an effective chiral theory obtained from a dimensionally reduced 
scalar field theory near the horizon of a Schwarzschild black hole \cite{iuw06,ms06}. Because
the computations are performed in a static background geometry, none of these approaches can 
address the processes that occur during gravitational collapse, neither can they take into account the 
effects of quantum gravity. 

In this paper we will examine the behavior of the wave functionals describing the quantized collapse 
of a self-gravitating dust cloud with a view to understanding the near horizon processes during gravitational
collapse. To do so we will use an exact canonical quantization of the LeMa\^\i tre-Tolman-Bondi (LTB) 
models \cite{ltb} that was developed not long ago by us in \cite{vws01,vws03,kmv05}.
The classical geometrodynamic constraints of the dust-gravity system are given in terms of a canonical 
chart consisting of the physical (area) radius of dust shells, the mass contained within a shell 
(the mass function), the dust proper time and their conjugate momenta. As usual, the system is described by 
a Hamiltonian constraint and a momentum constraint. For the LTB models it is possible to eliminate the 
momentum conjugate to the mass function using the momentum constraint and the resulting 
simplified Hamiltonian constraint is quadratic in the remaining momenta, yielding a Klein-Gordon-like 
functional differential equation, the Wheeler-DeWitt equation, when Dirac's quantization is applied. 
The quadratic Hamiltonian constraint requires regularization, which was performed on a lattice in 
\cite{kmv05}. When care is taken to ensure that the momentum constraint is satisfied in the continuum limit, 
the exact wave-functional describing quantum collapse becomes expressible in terms of lattice 
wave functions, each of which describes a collapsing shell and is determined by a set of three equations, 
{\it viz.,} the Hamilton-Jacobi equation and two additional constraints. These make it possible to 
determine unique, exact solutions everywhere, in the interior and in the exterior, as 
well as the Hilbert-space measure. 

We used these solutions to examine the post-collapse Hawking radiation spectrum by computing the 
Bogoliubov coefficient in the approximation in which a pre-existing black hole is surrounded by dust 
perturbations \cite{vksw02,kmsv07}. The mass function was taken to describe a central point mass $M$ plus 
the dust perturbations, which formally mimicked the quantum fields in the original Hawking derivation 
\cite{haw75}. In this calculation only the exterior wave-functional was necessary. Again, the Bekenstein-Hawking 
entropy \cite{bek72,bch73} of eternal black holes could also be determined via a microcanonical ensemble 
of states \cite{vw012,vgksw08,vw09}, but for this computation of the black hole entropy only the interior 
wave functional is necessary.

In the calculations mentioned in the previous paragraph no attempt was made to match the exterior and interior 
wave functionals or the shell wave functions at the apparent horizon during collapse. In this paper we will 
show that matching shell wave functions at the apparent horizon leads to an interesting picture, which requires 
an ingoing wave on one side 
of it to be necessarily accompanied by an outgoing wave on the other side. The relative amplitude of the 
outgoing wave is suppressed by the square root of the Boltzmann factor at a ``Hawking'' temperature that is 
inversely proportional to the mass contained within the shell.\footnote{Strictly speaking the Hawking temperature,
$T_H = (8\pi M)^{-1}$, refers to an eternal black hole of mass $M$. Here we show that the temperature 
appearing in the Boltzmann factor is $(4\pi F)^{-1}$, where $F$ is the mass function, which is a function
of the shell label and is equal to twice the mass contained within it. The temperature therefore varies from shell 
to shell, decreasing as one goes out from the center of the collapsing dust ball. We call it the ``Hawking''
temperature because of its similarity to the traditional Hawking temperature.}
Each pair, {\it i.e.,} an exterior ingoing wave plus an interior outgoing wave or an interior ingoing wave 
plus an exterior outgoing wave is an exact solution to the quantum mechanical problem, but neither is physically 
acceptable because neither conserves the flux of shells across the apparent horizon. To achieve shell flux 
conservation the two solutions must be suitably superposed so that, on either side of the apparent horizon,
the shell wave functions become the sum of an ingoing wave plus an outgoing wave whose amplitude is suppressed by the 
square root of the Boltzmann factor at the ``Hawking'' temperature. 

This can be viewed in one of two complementary ways. If one confines oneself to the exterior, ingoing matter 
appears to have a non-vanishing probability of being reflected at the horizon and the ratio of the reflection 
probability to the absorption probability is precisely the Boltzmann factor at the ``Hawking'' temperature for 
the shell. However, when the entire wave function is taken into account it becomes clear that a different 
point of view may also be adopted. As we will show, the outgoing modes in the interior have a unit relative 
probability of tunneling through the horizon but, because these modes are suppressed by the Boltzmann factor
relative to the ingoing modes in the exterior, the net effect is that the emission probability is still the 
Boltzmann factor at the Hawking temperature. This tunneling from interior to exterior may provide a 
mechanism by which information from the interior is transferred to the exterior.  

Diffeomorphism invariant wave functionals cannot be directly constructed from the shell wave functions 
described above. We discuss how they should be constructed and show how they too can be matched at 
the horizon. Just as the matching of shell wave functions leads to a new way to recover the Hawking 
temperature, matching the wave functionals provides a novel way to compute the Bekenstein-Hawking entropy of 
the end state black hole \cite{bek72,bch73}.

We review the essential features of the quantum LTB model and present the exact quantum states in section 
II. In section III we consider the behavior of shell wave functions across the horizon. Here we show that 
an incoming wave in one region must be accompanied by an outgoing wave in the other. We also argue that 
the two solutions must be superposed to conserve shell flux across the horizon. We discuss the construction 
of diffeomorphism invariant wave functionals in section IV. We show that the matching of wave-functionals 
implies that the ratio of the transmission probability to the absorption probability depends on the mass of 
the collapsing dust-ball and is given by the black hole Bekenstein-Hawking entropy. We 
close with a brief discussion in section V.

\section{Wavefunctionals of the Quantum LTB model} 

\subsection{LeMa\^\i tre-Tolman-Bondi Models}

The LTB models describe self-gravitating dust clouds of arbitrary matter distribution. The energy 
momentum tensor is $T_{\mu\nu} = \varepsilon(\tau,\rho) U_\mu U_\nu$ where $\varepsilon(\tau,\rho)$ 
is the dust energy density, $U^\mu(\tau,\rho)$ is the dust four velocity, $\tau$ is the dust proper time
and $\rho$ labels dust shells that form the dust cloud. The LTB line element reads
\beq
ds^2 = d\tau^2 - \frac{(\partial_\rho R)^2}{1+2E(\rho)} d\rho^2 - R^2(\tau,\rho) d\Omega^2
\label{ltbm}
\eeq
where $R(\tau,\rho)$ is the area radius and $E(\rho)$ is an arbitrary function of $\rho$ called the 
``energy function''. Einstein's equations lead to the expressions
\beq
\varepsilon(\tau,\rho) = \frac{\partial_\rho F}{R^2\partial_\rho R},~~ \partial_\tau R 
= -\sqrt{\frac FR + 2E}
\label{einstein}
\eeq
where $F(\rho)$ is yet another arbitrary function of the shell label coordinate $\rho$ called the 
mass function and the negative sign in the equation for $\partial_\tau R$ is required to describe 
collapse.

The mass function represents twice the weighted mass contained within the shell of label $\rho$. If 
a scaling is chosen so that the physical radius coincides with the shell label coordinates $\rho$ at 
$\tau=0$, then it can be expressed in terms of the energy density according to  
\beq
F(\rho) = \int_0^\rho \varepsilon(0,\rho')\rho'^2 d\rho'.
\eeq
The energy function, on the other hand, represents the total initial energy and depends on the 
velocity profile at the ``initial'' time, $v(\rho)=[\partial_\tau R(\tau,\rho)]_{\tau=0}$,
\beq
2E(\rho) = v^2(\rho) - \frac 1\rho\int_0^\rho \varepsilon(0,\rho')\rho'^2d\rho'
\eeq
In this paper we shall concentrate only on the marginal models, for which $E(\rho)=0$. In this 
case, the solution to \eqref{einstein} is
\beq
R(\tau,\rho) = \rho\left[ 1-\frac 32\sqrt{\frac{F(\rho)}{\rho^3}}~\tau\right]^{2/3}.
\eeq
The central singularity is defined by the curve
\beq
\tau(\rho) = \frac{2\rho^{3/2}}{3\sqrt{F(\rho)}},
\eeq
which gives the proper time at which shells achieve zero physical radius. Of course, so long as
the above equation for $\tau$ is not satisfied, $\rho=0$, which is also gives $R=0$ by our choice of 
scaling, is not a physical singularity. Various models are obtained from choices of $F(\rho)$: even 
the eternal black hole can be described by the LTB solution in 
\eqref{ltbm} by choosing $F(\rho)=2M\Theta(\rho)$, where $\Theta$ is the unit step function.

\subsection{Quantization}

The general spherically symmetric Arnowitt-Deser-Misner (ADM) metric,
\beq
ds^2 = N^2 dt^2 - L^2(dr-N^r dt)^2 -R^2d\Omega^2
\eeq
can be embedded in the spacetime described by \eqref{ltbm}. After a series of canonical 
transformations described in detail in \cite{vws01} and \cite{kmv05}, this procedure leads to a 
canonical description of the classical black hole in terms of the dust proper time, $\tau(r)$, 
the area radius $R(r)$, the mass density function $\Gamma(r)$ defined by
\beq
F(r) = \int_0^r dr' \Gamma(r'),
\eeq
and their conjugate momenta, $P_\tau(r)$, $P_R(r)$ and $P_\Gamma(r)$ respectively. The constraints 
of the self-gravitating dust system in these variables can then be given as
\bea
H &=& P_\tau^2 +\cF P_R^2 -\frac{\Gamma^2}{4\cF} \approx 0\cr\cr
H_r &=& \tau' P_\tau + R'P_R -\Gamma P_\Gamma' \approx 0
\label{const}
\eea
where $\cF = 1-F/R$.

Dirac's procedure may be employed to quantize this system. The momenta must be replaced by
functional derivatives with respect to their corresponding configuration variables. There are,
however, regularization and factor ordering ambiguities that can be encapsulated at a formal 
level by introducing factors of $\delta(0)$ into the functional Schroedinger equation and 
writing the Hamiltonian constraint as \cite{kmv05}
\beq
\widehat H \Psi[\tau,R,\Gamma] = \left[\frac{\delta^2}{\delta\tau(r)^2} + \cF \frac{\delta^2}
{\delta R(r)^2} + A \delta(0) \frac{\delta}{\delta R(r)} + B \delta(0)^2 + \frac{\Gamma^2}
{4\cF}\right]\Psi[\tau,R,\Gamma] \approx 0
\label{wdeqn}
\eeq
where $A(R,F)$ and $B(R,F)$ are arbitrary functions of $R,F$, which serve to take into account
the factor ordering ambiguities. The divergent factors $\delta(0)$ indicate that the factor 
ordering problem is unsolved and can be dealt with only once a suitable regularization 
of the Wheeler-DeWitt equation has been performed. The second (diffeomorphism) constraint in
\eqref{const} requires no regularization and may be written as
\beq
\left[\tau'\frac{\delta}{\delta \tau(r)}+R'\frac{\delta}{\delta R(r)} - \Gamma \left(
\frac{\delta}{\delta \Gamma}\right)'\right]\Psi[\tau,R,\Gamma] \approx 0
\eeq
Progress with the constraints can only be made after a suitable regularization has been 
applied. Below we briefly describe and apply a lattice regularization.

\subsection{Lattice Regularization}

For solutions of the constraints, we make the ansatz
\beq
\Psi[\tau,R,\Gamma] = \Psi^{(0)}[F] \exp\left[-\frac i 2 \int dr~ \Gamma(r)~ \cW(\tau,R,F)\right],
\label{ansatz}
\eeq
where $\cW(\tau,R,F)$ is some function to be determined. It automatically satisfies the diffeomorphism 
constraint. The Wheeler-DeWitt equation is second order in time derivatives so both positive and negative 
energy solutions exist, but we will confine our attention to the positive energy solutions above. It is 
worth noting that any functional
\beq
\Psi[\tau,R,\Gamma] = U\left(-\frac i 2 \int dr~ \Gamma(r)~ \cW(\tau,R,F)\right)
\label{genform}
\eeq
would satisfy the diffeomorphism constraint provided that $\cW$ has no explicit dependence on 
the label coordinate $r$ except through the mass function, $F(r)$. We have chosen $U=\exp$
so that the wave-functional may also be factorizable on a spatial lattice, whose cell size we
call $\sigma$, taking $\sigma\rightarrow 0$ in the continuum limit. Diffeomorphism invariance 
requires that the continuum wave-functional and all physical results be independent of the cell 
size. On the lattice, the argument of the exponential function becomes \cite{vws03,kmv05}
\beq
\int dr~ \Gamma(r)~ \cW(\tau,R,F) \rightarrow \sigma \sum_j \Gamma_j \cW(\tau_j,R_j,F_j)
\eeq
where $\Gamma_j = \Gamma(r_j)$, etc. This turns the wave-functional into a product state,
\beq
\Psi[\tau,R,\Gamma] = \prod_j \psi_j(\tau_j,R_j,F_j) = \prod_j \psi^{(0)}_j \exp\left[-\frac i2 \sigma 
\sum_j \Gamma_j \cW(\tau_j,R_j,F_j)\right]
\label{solnform}
\eeq
provided that $U=\exp$.

Before proceeding further it is necessary to define what is meant by a functional derivative when 
functions are defined on a lattice \cite{vws03}. The defining equations can be understood by analogy with the 
simplest properties of functional derivatives of the functions $J(x)$
\bea
&&\frac{\delta J(y)}{\delta J(x)} = \delta(y-x),\cr\cr
&&\frac{\delta}{\delta J(x)}\int dyJ(y)  = 1
\eea
and from these definitions follows
\beq
\frac{\delta}{\delta J(x)}\int dyJ(y)\phi(y)=\phi(x).
\eeq
On a lattice we define, for the lattice intervals $x_i$ and $x_j$,
\beq
\frac{\delta J(x_{i})}{\delta J(x_{j})}=\Delta(x_i-x_j) = \lim_{\sigma \rightarrow 0}
\frac{\delta_{ij}}\sigma
\label{funder}
\eeq
where $r_i$ labels the $i^\text{th}$ lattice site and $\delta_{ij}$ is the Kronecker $\delta$, equal 
to zero when the lattice sites $x_i$ and $x_j$ are different and one when they are the same. 
Just as $\delta(y-x)$ is only defined as an integrand in an integral, so $\Delta(x_{i}-x_{j})$ 
should also be considered defined only as a summand in a sum over lattice sites. Hence
\beq
\lim_{\sigma\rightarrow 0} \frac{\delta}{\delta J(r_j)} \sigma\sum_i J(r_i) = 
\lim_{\sigma \rightarrow 0}\sigma\sum_i \frac{\delta J(r_i)}{\delta J(r_j)}=1
\eeq
and
\beq
\frac{\delta}{\delta J(r_j)} \sigma\sum_i  J(r_i)\phi(r_i) = \lim_{\sigma\rightarrow 0} \sigma \sum_i 
\Delta(r_i-r_j)\phi(r_i) = \phi(r_j)
\eeq
It follows that
\beq
\frac{\delta}{\delta J(x_{j})}\rightarrow \frac 1\sigma\lim_{\sigma \rightarrow 0} \frac{\partial}
{\partial J_j}
\label{funder2}
\eeq
where $J_j=J(x_j)$. This is compatible with the formal (continuum) definition of the
functional derivative. 

\subsection{Collapse Wave Functionals}

When \eqref{funder} and \eqref{funder2} are applied to the Wheeler-DeWitt equation in \eqref{wdeqn}
and $\Psi[\tau,R,\Gamma]$ is taken to be a product state, one obtains an equation describing the wave 
functions at each lattice point \cite{kmv05}
\beq
\left[\frac{\partial^2}{\partial\tau_j^2} + \cF_j \frac{\partial^2}{\partial R_j^2} + A_j 
\frac{\partial}{\partial R_j} + B _j + \frac{\sigma^2\Gamma_j^2}{4\cF_j}\right]\psi_j \approx 0,
\label{wd1}
\eeq
but there is a further restriction arising from the diffeomorphism constraint. Inserting the ansatz 
in \eqref{solnform} into \eqref{wd1}, we find
\bea
&&\frac{\sigma^2\Gamma_j^2}4 \left[\left(\frac{\partial \cW_j}{\partial \tau_j}\right)^2 + \cF_j
\left(\frac{\partial \cW_j}{\partial R_j}\right)^2 - \frac 1\cF\right]\cr\cr
&&\hskip 2cm \frac{\sigma\Gamma_j}2\left[\frac{\partial^2 \cW_j}{\partial \tau_j^2} + 
\cF_j\frac{\partial^2 \cW_j}{\partial R_j^2} + A_j \frac{\partial \cW_j}{\partial R_j}\right] + B_j = 0,
\eea
which must be satisfied {\it independently} of $\sigma$. This is only possible if the following 
three equations are simultaneously satisfied at each lattice site \cite{kmv05},
\bea
&&\left[\left(\frac{\partial \cW_j}{\partial\tau_j}\right)^2 + \cF_j 
\left(\frac{\partial \cW_j}{\partial R_j}\right)^2 - \frac{1}{\cF_j}\right]=0,\cr\cr
&&\left[\frac{\partial^2 \cW_j}{\partial\tau_j^2} + \cF_j\frac{\partial^2 \cW_j}
{\partial R_j^2} + A_j \frac{\partial \cW_j}{\partial R_j} \right]  =  0,\cr\cr
&& B_j = 0.
\label{4eqns}
\eea
Moreover, it is straightforward that the Hamiltonian constraint is Hermitean if and only if 
\beq
A_j = \cF_j\partial_{R_j} \ln ({\mathfrak m}_j |\cF_j|).
\eeq
where $\mathfrak m_j$ is the Hilbert space measure.

Unique solutions to the equations in \eqref{4eqns} and having the form given in \eqref{solnform} 
have been obtained in all, even the non-marginally bound, cases \cite{kmv05}. For the marginally 
bound models the solution for the phase $\cW_j$ in the exterior, {\it i.e.,} for shells that lie 
outside the apparent horizon ($R_j>F_j$), is
\beq
\cW_j^{(\pm)} = \tau_j \pm 2F_j\left[z_j -\tanh^{-1}\frac 1{z_j}\right],~~ z_j > 1
\eeq
where $z_j=\sqrt{R_j/F_j}$. The positive sign refers to ingoing waves, traveling toward the horizon and 
the negative sign to outgoing waves, as can be seen from the signature of the phase velocity,
\beq
\dot z_j = \mp \frac{z_j^2-1}{2F_jz_j^2},
\eeq
keeping in mind that $z_j>1$. In the interior,  {\it i.e.,} for shells that lie inside the apparent horizon  
($R_j<F_j$), the solution is 
\beq
\cW_j^{(\pm)} = \tau_j \pm 2F_j\left[z_j -\tanh^{-1} z_j \right],~~ z_j < 1
\eeq
but here the the positive sign refers to outgoing waves and the negative sign to ingoing waves, traveling 
toward the central singularity, again as determined by the phase velocity. Furthermore, 
as shown in Appendix B of \cite{kmv05},  the system in \eqref{4eqns} determines not only  $\cW_j$ but 
the Hilbert space measure, $\mathfrak{m}_j$, as well. For the marginal models under consideration, 
$\mathfrak{m_j}$ is regular everywhere and given by
\beq
\mathfrak m_j = z_j
\eeq
upto a constant scaling.

\section{Shell Wave Functions}

The interior and exterior wave functions must be matched at the apparent horizon, but the phases diverge 
there because the apparent horizon is an essential singularity of \eqref{4eqns}. We can perform the matching, 
however, if we analytically continue to the complex plane and consider the functions
\beq
\cW^{(\pm)}(\tau_j,z_j)=\left\{\begin{matrix}
\tau_j \pm 2F_j\left[z_j -\acoth z_j\right], & \cr\cr
\tau_j \pm 2F_j\left[z_j -\atanh z_j\right], & 
 \end{matrix}\right.
 \eeq
where $\atanh(z)$ and $\acoth(z)$ refer respectively to the principal value of the inverse hyperbolic 
tangent and cotangent functions, for which the following identities are well known \cite{abramstegun}:
\beq
\acoth z = \atanh \frac 1z
\eeq
and
\beq
\acoth z = \atanh z + \left\{\begin{matrix}
\frac{i\pi}2 & \text{Im}(z) \leq 0\cr\cr
-\frac{i\pi}2 & \text{Im}(z) > 0
\end{matrix}\right.
\label{matchprop}
\eeq
The functions 
\beq
\cW^{(\pm)}_\text{out}(\tau_j,z_j)= \tau_j \pm 2F_j\left[z_j -\acoth z_j\right],
\eeq
which refer to waves in the region outside the apparent horizon, can be transformed into the functions 
\beq
\cW^{(\pm)}_\text{in}(\tau_j,z_j)= \tau_j \pm 2F_j\left[z_j -\atanh z_j\right],
\eeq
which refer to waves in the region inside, by employing the property \eqref{matchprop} above 
for $\text{Im}(z)\leq 0$. First, assuming that the proper time of the dust shell, $\tau_j$, remains 
unaffected by the transformation, one finds
\beq
\cW^{(+)}_\text{out}(\tau_j,z_j)= \cW^{(+)}_\text{in}(\tau_j,z_j) - i\pi F_j
\eeq 
 and
 \beq
\cW^{(-)}_\text{out}(\tau_j,z_j)= \cW^{(-)}_\text{in}(\tau_j,z_j) + i\pi F_j.
\eeq 
Thus an ingoing wave of positive energy in the exterior gets transformed into an outgoing wave in 
the interior according to 
\beq
\psi_\text{out}^{(+)}(\tau_j,R_j,F_j) = e^{-\pi\omega_jF_j} \psi_\text{in}^{(+)}(\tau_j,R_j,F_j)
\eeq
where $\omega_j = \sigma\Gamma_j/2$ is the shell energy. Likewise an ingoing wave of positive 
energy in the interior is transformed into an outgoing wave in the exterior
\beq
\psi_\text{in}^{(-)}(\tau_j,R_j,F_j) = e^{-\pi\omega_jF_j} \psi_\text{out}^{(-)}(\tau_j,R_j,F_j)
\eeq
One can easily check that the derivatives of the states coincide in the lower half plane as well. 

We can now give two separate solutions to the quantum mechanical problem: since the exterior, 
ingoing wave is matched only to an interior, outgoing wave,
\beq
\psi^{(1)}_j = \left\{\begin{matrix}
e^{-i\omega_j\left[\tau_j + 2F_j \left(\sqrt{\frac{R_j}{F_j}} -\tanh^{-1}\sqrt{\frac{F_j}{R_j}}\right)\right]} 
& R_j>F_j\cr\cr
e^{-\pi\omega_jF_j} e^{-i\omega_j\left[\tau_j + 2F_j \left(\sqrt{\frac{R_j}{F_j}} -\tanh^{-1}
\sqrt{\frac{R_j}{F_j}}\right)\right]} & R_j<F_j
\end{matrix}\right.
\label{wf1}
\eeq
will be continuous and differentiable everywhere. Likewise the interior, ingoing wave is matched only 
to an exterior, outgoing wave, therefore
\beq
\psi^{(2)}_j = \left\{\begin{matrix}
e^{-\pi\omega_jF_j} e^{-i\omega_j\left[\tau_j - 2F_j \left(\sqrt{\frac{R_j}{F_j}} -\tanh^{-1}
\sqrt{\frac{F_j}{R_j}}\right)\right]}& R_j>F_j\cr\cr
e^{-i\omega_j\left(\tau_j - 2F_j \left[\sqrt{\frac{R_j}{F_j}} -\tanh^{-1}\sqrt{\frac{R_j}{F_j}}\right)\right]} 
& R_j<F_j
\end{matrix}\right.
\label{wf2}
\eeq
is also continuous and differentiable everywhere. The problem is that neither of these solutions is
physically acceptable: the wave functions in \eqref{wf1} represent a flow toward the apparent horizon both in the 
exterior as well as in the interior, while the wave functions in \eqref{wf2} represent a flow
away from the apparent horizon, again in both regions.  Therefore we consider a linear superposition of the two solutions,
\beq
\psi_j = \left\{\begin{matrix}
e^{-i\omega_j\left[\tau_j + 2F_j \left(\sqrt{\frac{R_j}{F_j}} -\tanh^{-1}\sqrt{\frac{F_j}{R_j}}\right)
\right]} + A_j e^{-\pi\omega_jF_j} e^{-i\omega_j\left[\tau_j - 2F_j \left(\sqrt{\frac{R_j}{F_j}} -\tanh^{-1}
\sqrt{\frac{F_j}{R_j}}\right)\right]}& R_j>F_j\cr\cr
e^{-\pi\omega_jF_j} e^{-i\omega_j\left[\tau_j + 2F_j \left(\sqrt{\frac{R_j}{F_j}} -\tanh^{-1}
\sqrt{\frac{R_j}{F_j}}\right)\right]} + A_j e^{-i\omega_j\left[\tau_j - 2F_j \left(\sqrt{\frac{R_j}{F_j}} -
\tanh^{-1}\sqrt{\frac{R_j}{F_j}}\right)\right]} & R_j<F_j
\end{matrix}\right.
\label{wavefns0}
\eeq
where $A_j$ are constants to be determined. 

From the Klein-Gordon-like equation \eqref{wd1} for shells one obtains a continuity equation, 
$\partial_\mu J^\mu_j =0$, in which we define 
\beq
J_j^\tau =- i \text{sgn}(\cF)\mathfrak{m_j} \psi^*_j\overleftrightarrow{\partial_{\tau_j}} \psi_j,~~ 
J^R_j = -i {\mathfrak m}_j |\cF_j| \psi^*_j \overleftrightarrow{\partial_{R_j}} \psi_j,
\eeq
where the components of $J^\mu_j$ have been chosen so that the shell current density, $J^R_j$, is 
continuous across the horizon. We cannot give an interpretation for $J^\tau_j$ within this 
quantum mechanical model, but since $J^\tau_j$ is time independent it follows that $J^R_j$ is 
divergence free, therefore constant. In the exterior and in the interior we find 
\beq
J^R_j = \left\{ \begin{matrix}
2\omega (|A_j|^2 e^{-2\pi \omega_j F_j}-1), & R_j >F_j\cr\cr
2\omega (e^{-2\pi \omega_j F_j}-|A_j|^2), & R_j <F_j
\end{matrix}\right.
\eeq
and so $|A_j|^2=1$. Therefore, we take $A_j=1$ for every shell. This gives the absorption 
probability, $P_\text{abs,j}=1$, which is reasonable since there is no barrier for a shell to cross the 
apparent horizon from the exterior. The right hand side of the first expression in \eqref{wavefns0} is 
the exterior wave function of a shell. Its first term is the ingoing wave that 
represents the collapsing shell. Its second term is an outgoing wave that represents a shell reflection 
at the horizon and an external observer could view the factor $R_\text{ext} = e^{-\pi\omega_j F_j}$ in 
$\psi_\text{out}$ as the reflection coefficient \cite{k04,kf04}. From such an observer's perspective an 
ingoing wave is necessarily accompanied by a reflected wave at the apparent horizon and the ratio of 
the reflection probability to the absorption probability is the Boltzmann factor
\beq
\frac{P_\text{ref,j}}{P_\text{abs,j}} = e^{-2\pi \omega_j F_j},
\eeq
at the temperature 
\beq
T_F = (2\pi F_j)^{-1}.
\eeq
This is twice the ``Hawking'' temperature.  A similar discrepancy between the calculated temperature and the Hawking 
temperature was also noted in early analyses of the canonically invariant, null geodesic tunneling formula
\cite{c06,m06,aas06,p08,cc09}.  It was later discovered \cite{n07,apgs08,apgs09} that the discrepancy arose because an 
additional contribution, coming from the temporal part, had been ignored.  We will now show that the same is true in 
our approach.

Let  us see how an additional contribution from the temporal part comes about in our picture by reconsidering our 
assumption that the shell proper time remains unaffected by the analytic continuation. In the marginal models the 
proper time is related to the momentum, $P_\Gamma$, conjugate to the mass density, $\Gamma$, by \cite{kmv05}
\beq
\tau = 2P_\Gamma \pm 2F \left[z -\acoth z\right]
\eeq
in the exterior and
\beq
\tau = 2P_\Gamma \pm 2F \left[z -\atanh z\right]
\eeq
in the interior, where the positive sign in both cases is for ingoing matter and the negative sign for 
outgoing matter. If $P_\Gamma$ (and not $\tau$) remains unchanged by the continuation then, analytically 
continuing as before in the lower half plane, we find
\beq
\tau_\text{out} = \tau_\text{in} \mp i \pi F.
\label{ptimecont}
\eeq
In the standard computations leading to the Hawking temperature via the Bogoliubov 
coefficient, modes are defined with respect to Killing time and then compared at $\Im^+$. Although, in the 
midst of the collapsing dust ball there is no time-like Killing vector and therefore no Killing time, 
$P_\Gamma$ has the interpretation of one half the Killing time in the Schwarzschild spacetime outside 
collapsing ball and therefore is its natural replacement inside. 

The additional contribution from the proper time in \eqref{ptimecont} implies that
\bea
&&\cW^{(+)}_\text{out}(\tau_j,z_j)= \cW^{(+)}_\text{in}(\tau_j,z_j) - 2i\pi F\cr\cr
&&\cW^{(-)}_\text{out}(\tau_j,z_j)= \cW^{(-)}_\text{in}(\tau_j,z_j) + 2i\pi F
\eea
so that \eqref{wf1} and \eqref{wf2} should read respectively,
\beq
\psi^{(1)}_j = \left\{\begin{matrix}
e^{-i\omega_j\left[\tau_j + 2F_j \left(\sqrt{\frac{R_j}{F_j}} -\tanh^{-1}\sqrt{\frac{F_j}{R_j}}\right)\right]} 
& R_j>F_j\cr\cr
e^{-2\pi\omega_jF_j} e^{-i\omega_j\left[\tau_j + 2F_j \left(\sqrt{\frac{R_j}{F_j}} -\tanh^{-1}
\sqrt{\frac{R_j}{F_j}}\right)\right]} & R_j<F_j
\end{matrix}\right.
\label{wf3}
\eeq
and 
\beq
\psi^{(2)}_j = \left\{\begin{matrix}
e^{-2\pi\omega_jF_j} e^{-i\omega_j\left[\tau_j - 2F_j \left(\sqrt{\frac{R_j}{F_j}} -\tanh^{-1}
\sqrt{\frac{F_j}{R_j}}\right)\right]}& R_j>F_j\cr\cr
e^{-i\omega_j\left(\tau_j - 2F_j \left[\sqrt{\frac{R_j}{F_j}} -\tanh^{-1}\sqrt{\frac{R_j}{F_j}}\right)\right]} 
& R_j<F_j
\end{matrix}\right. 
\label{wf4}
\eeq
From these we construct the (corrected) shell wave functions by superposition
\beq
\psi_j = \left\{\begin{matrix}
e^{-i\omega_j\left[\tau_j + 2F_j \left(\sqrt{\frac{R_j}{F_j}} -\tanh^{-1}\sqrt{\frac{F_j}{R_j}}\right)\right]} 
+ e^{-2\pi\omega_jF_j} e^{-i\omega_j\left[\tau_j - 2F_j \left(\sqrt{\frac{R_j}{F_j}} -\tanh^{-1}
\sqrt{\frac{F_j}{R_j}}\right)\right]}& R_j>F_j\cr\cr
e^{-2\pi\omega_jF_j} e^{-i\omega_j\left[\tau_j + 2F_j \left(\sqrt{\frac{R_j}{F_j}} -\tanh^{-1}
\sqrt{\frac{R_j}{F_j}}\right)\right]} + e^{-i\omega_j\left[\tau_j - 2F_j \left(\sqrt{\frac{R_j}{F_j}} -
\tanh^{-1}\sqrt{\frac{R_j}{F_j}}\right)\right]} & R_j<F_j
\end{matrix}\right.
\label{wavefns}
\eeq
The ratio of the reflection probability to the absorption probability is now the Boltzmann factor at the 
``Hawking'' temperature appropriate for the shell. From the point of view of the collapse, this ``reflected'' 
piece of the external wave function is a purely quantum effect necessitated by the existence of an ingoing 
wave in the interior. This is not the same as the tunneling picture of black hole evaporation in which pair 
production occurring near the horizon causes a negative energy antiparticle to fall into the black hole 
and a positive energy particle to tunnel through into the exterior. Here we have considered only positive 
energy solutions. 

The right hand side of the second expression in \eqref{wavefns} is the interior wave function of a shell. 
Its second term is an ingoing wave that represents a shell continuing its collapse into the central 
singularity. Without recourse to boundary conditions at the center, which would necessarily be {\it ad hoc}, 
we see that it must be accompanied by an outgoing wave (the first term in the same expression). This 
outgoing wave comes with amplitude $R_\text{int} = e^{-2\pi\omega_j F_j}$, precisely equal to the amplitude 
for reflection on the horizon. Once again this is a purely quantum effect that is required by the existence 
of an ingoing wave in the exterior. 

\section{Collapse Wave Functionals}

The manner in which the shell wave functions match across the horizon is essential for the construction 
of diffeomorphism invariant wave functionals describing the collapse as we argue in the following section.
The superposed wave functions in \eqref{wavefns} cannot be used to construct diffeomorphism invariant 
wave functionals describing the collapse, since they are not simple exponentials as required 
by \eqref{genform}. Nevertheless, linear superpositions of diffeomorphism invariant functionals will be 
diffeomorphism invariant and we can construct diffeomorphism invariant functionals from each of \eqref{wf3} 
and \eqref{wf4} by taking the continuum limit of the product over shells. Thus in the continuum limit \eqref{wf3} 
turns into
\beq
\Psi_1 = \left\{\begin{matrix}
e^{-\frac i2\int dr~ \Gamma~ \cW^{(+)}_\text{out}(\tau,R,F)}  & R>F\cr\cr
e^{-\pi \int dr \Gamma F} e^{ -\frac i2 \int dr~ \Gamma~ \cW^{(+)}_\text{in}(\tau,R,F)}& R<F
\end{matrix}\right.
\eeq
whereas \eqref{wf4} into
\beq
\Psi_2 = \left\{\begin{matrix}
e^{-\pi \int dr \Gamma F} e^{-\frac i2 \int dr~ \Gamma~ \cW^{(-)}_\text{out}(\tau,R,F)} & R>F\cr\cr
e^{-\frac i2 \int dr~ \Gamma~ \cW^{(-)}_\text{in}(\tau,R,F)} & R < F
\end{matrix}\right.
\eeq
and because $\Gamma=F'(r)$, the amplitude in each case is
\beq
\int_0^\infty dr \Gamma(r) F(r)  = \frac 12(F^2(\infty)-F^2(0)) = 2M^2,
\eeq
where $M$ is the ADM mass of the collapsing dust-ball. Thus the diffeomorphism invariant functional 
equivalent of the superposed shell wave functions in \eqref{wavefns} is $\Psi=\Psi_1+\Psi_2$ or
\beq
\Psi = \left\{\begin{matrix}
e^{-\frac i2\int dr~ \Gamma~ \cW^{(+)}_\text{out}(\tau,R,F)} + e^{-2\pi M^2} e^{-\frac i2 
\int dr~ \Gamma~ \cW^{(-)}_\text{out}(\tau,R,F)} & R >F \cr\cr
e^{-2\pi M^2} e^{ -\frac i2 \int dr~ \Gamma~ \cW^{(+)}_\text{in}(\tau,R,F)}
+ e^{-\frac i2 \int dr~ \Gamma~ \cW^{(-)}_\text{in}(\tau,R,F)}& R <F
\end{matrix}\right.
\eeq
and the ratio of the reflection probability to the probability for absorption is 
\beq
\frac{P_\text{ref}}{P_\text{abs}} = e^{-4\pi M^2} \equiv e^{-S},
\label{entropy}
\eeq
where $S$ is the entropy of the end state black hole. A similar matching of the wave functionals 
was also found in \cite{kmv05}. This provides an alternative and direct 
derivation of the Benkenstein-Hawking entropy in the same way as the matching of shell wave 
functions provides a derivation of the Hawking temperature. However, the extent to which this
result is generic remains to be tested.

\section{Discussion}

Let us first summarize the main results of this paper. We have used a canonical quantization of 
the marginal LeMa\^\i tre-Tolman-Bondi model of a self-gravitating dust ball to understand the 
near horizon behavior of the wave-functionals describing collapse. There is generally a factor 
ordering ambiguity in any attempt to quantize gravity, which can be resolved only after a suitable
regularization has been chosen. As in previous works, here regularization was performed
on a lattice. With this regularization exact solutions of the Wheeler-DeWitt were obtained. They 
cannot be directly matched at the horizon, which is an essential singularity of the Wheeler-DeWitt
equation. Beginning with the shell wave functions, we performed an analytic continuation into the 
complex plane and showed that matching each shell's wave function at the horizon requires that an 
ingoing wave function in one region is accompanied by an outgoing wave function in the other. 
The amplitude of the outgoing wave function in each case is given by the square root of the Boltzmann 
factor at the ``Hawking'' temperature defined by the mass contained within that shell. Furthermore the 
two solutions describing an ingoing wave in one region and an exponentially suppressed outgoing 
wave in the other region must be superposed to conserve the shell flux across the 
horizon. We showed that crossing the horizon involved a rotation of the dust proper time. This 
rotation is required to keep $P_\Gamma$, which is related to the Killing time in the exterior, 
unchanged. 

The probability for an ingoing shell to cross the horizon is unity. There is also an outgoing 
shell in the exterior, which we have called a ``reflected'' shell but it can also be viewed as 
a tunneling of an outgoing wave in the interior into the exterior with unit relative probability.
The ratio $P_\text{ref}/P_\text{abs} = e^{-\beta_H \omega}$, where $\omega$ is the infalling shell's energy 
and $\beta_H$ is the inverse ``Hawking'' temperature. Matching wave-functionals across the horizon 
yields the same picture, but now $P_\text{ref}/P_\text{abs} = e^{-S}$, where $S$ is the entropy of 
the black hole that is the end state of the collapsing dust.

The lattice regularization scheme we have used correctly implements diffeomorphism invariance in the 
continuum limit. It also uniquely fixes the factor ordering and it is possible to obtain exact 
and unique solutions to all the constraints. Thus these are the only solutions in which states 
for the dust cloud factorize into (infinitely many) shell states. Other solutions would couple 
the shells, but to find them one would have to find a regularization scheme that is different 
from the one chosen here. The factorizable solutions are, however, just the WKB solutions, 
{\it i.e.,} the regularization and resulting factor ordering used here lead to states for which 
the WKB form is exact. If a different regularization scheme, consistent with diffeomorphism 
invariance, can be found then it will produce corrections to the solutions given here, in particular 
to the amplitudes for crossing the horizon and to the black hole entropy as computed via \eqref{entropy}. 

Nevertheless the techniques used here should still be applicable. Let us close by illustrating the 
effect of corrections using the results of a recent paper \cite{bkm10} in which the solutions have 
been given up to order $l_p^2$, where $l_p$ is the Planck length. This will also show off the great 
simplicity of this approach over the traditional Bogoliubov calculation of the Planck spectrum. 
Interpreting the solutions given in \cite{bkm10} as shell wave functions (extended to the complex 
plane), we have 
\beq
\cW^{(\pm)}(\tau_j,z_j)=\left\{\begin{matrix}
(1+Cl_p^2)\tau_j \pm 2F_j\left[z_j + \frac 13 C l_p^2 z^3 -\acoth z_j\right], \text{Re}(z)>1& \cr\cr
(1+Cl_p^2)\tau_j \pm 2F_j\left[z_j + \frac 13 C l_p^2 z^3 -\atanh z_j\right], \text{Re}(z)<1&
 \end{matrix}\right.
 \eeq
where $C$ is a constant. Matching now picks up an additional multiple of $\mp i\pi F$ coming from the 
multiplier of $\tau_j$. Thus we find
\bea
&&\cW^{(+)}_\text{out}(\tau_j,z_j)= \cW^{(+)}_\text{in}(\tau_j,z_j) - 2i\pi F(1+\frac 12 Cl_p^2)\cr\cr
&&\cW^{(-)}_\text{out}(\tau_j,z_j)= \cW^{(-)}_\text{in}(\tau_j,z_j) + 2i\pi F(1+\frac 12 Cl_p^2)
\eea
This gives the ratio of the reflection probability to the absorption probability as
\beq
\frac{P_\text{ref}}{P_\text{abs}} = e^{-4\pi \omega_j F_j (1+\frac 12 C l_p^2)}
\eeq
which is the Boltzmann factor at the modified ``Hawking'' temperature
\beq
T_H = \frac 1{4\pi F_j (1+\frac 12 C l_p^2)}
\eeq
Likewise the corrections to the Bekenstein-Hawking entropy are obtained by matching the wave-functionals 
at the horizon as we have done earlier:
\beq
e^{-S} = e^{-2\pi (1+\frac C2 l_p^2)\int_0^\infty dr \Gamma F} = e^{-4\pi M^2 (1+\frac C2 l_p^2)}
\eeq
In this paper, we have only examined the marginal models for the purpose of illustration. The general 
solutions for non-marginal models have also been worked out both within the lattice regularization 
described in our earlier papers as well as by the authors in \cite{bkm10}. We intend to examine them 
within the framework described here in a future publication.
\bigskip\bigskip

\noindent{\bf Acknowledgements}
\bigskip

\noindent LCR Wijewardhana was supported in part by the U.S. Department of Energy Grant No. 
DE-FG02-84ER40153.

\end{document}